\documentclass[%
twocolumn,
 amsmath,amssymb,
 aps,
 prl,
]{revtex4}

\usepackage{graphicx}
\usepackage{dcolumn}
\usepackage{bm}
\usepackage{graphicx}
\usepackage{braket}
\usepackage{float}
\usepackage{natbib}
\usepackage{subfigure}
\usepackage{color}
\usepackage[mathlines]{lineno}
\usepackage{verbatim} 
\usepackage{soul}
\begin{document}


\title{Electron-phonon coupling in a magic-angle twisted-bilayer graphene device}

\author{Andreij C. Gadelha$^{1,2,\dag\dag}$}
\author{Viet-Hung Nguyen$^{3,\dag\dag}$}
\author{Eliel G. Neto$^{4}$}
\author{Fabiano Santana$^{1}$}
\author{Markus B. Raschke$^{2}$}
\author{Michael Lamparski$^{5}$}
\author{Vincent Meunier$^{5}$}
\author{Jean-Christophe Charlier$^{3}$}
\email{jean-christophe.charlier@uclouvain.be}
\author{Ado Jorio$^{1}$}
\email{adojorio@fisica.ufmg.br}

\affiliation{$^{1}$Physics Department, Universidade Federal de Minas Gerais, Belo Horizonte, MG, 31270-901, Brazil.}
\affiliation{$^{2}$Department of Physics, and JILA, University of Colorado at Boulder, Boulder, Colorado 80309,
USA.}
\affiliation{$^{3}$Institute of Condensed Matter and Nanosciences, Université catholique de Louvain (UCLouvain), Louvain-la-Neuve, Belgium.}
\affiliation{$^{4}$Physics Institute, Universidade Federal da Bahia, Salvador - BA, 40170-115 Brazil.}
\affiliation{$^{5}$Department of Physics, Applied Physics, and Astronomy, Jonsson Rowland Science 
Center, Troy, NY, USA.}
\affiliation{$^{\dag\dag}$ These two authors contributed equally to this work.}

\date{\today}

\begin{abstract}
The importance of phonons in the strong correlation phenomena observed in twisted bilayer graphene (TBG) at the so-called magic-angle is under debate.  Here we apply gate-dependent micro-Raman spectroscopy to monitor the G band linewidth in TBG devices of twist angles $\theta=0^{\circ}$, $\sim 1.1^{\circ}$ (magic-angle) and $\sim 7^{\circ}$ (large angle). The results show a broad and p/n-asymmetric doping behavior at the magic-angle, in clear contrast to the behavior observed in twist angles above and below. Atomistic modeling reproduces the experimental observations, revealing how the unique electronic structure of magic-angle TBGs influences the electron-phonon coupling and, consequently, the G band linewidth. Our findings indicate a connection between electron-phonon coupling and experimental observations of strongly correlated phenomena in magic-angle TBG. 
\end{abstract}
\maketitle


The unusual electronic structure of graphene, with massless Dirac fermions, along with the relatively light C atoms bound together by strong covalent $\sigma$ bonds is responsible for electron-phonon scattering that rules the lifetime of phonons, in departure from the broadly adopted adiabatic Born--Oppenheimer approximation \cite{pisana2007}. The breakdown fo the adiabatic Born--Oppenheimer approximation generates a Kohn anomaly \cite{lazzeri2006nonadiabatic}, which has been studied in graphite \cite{bonini2007}, single-layer graphene \cite{pisana2007,bonini2007,lazzeri2006nonadiabatic,hasdeo2016fermi}, double-layer graphene with the graphite-like AB (Bernal) stacking \cite{yan2008observation,mafra2009observation,das2009phonon,mafra2012characterizing}, and in metallic carbon nanotubes \cite{lazzeri2006phonon,tsang2007doping,farhat2007phonon} using Raman spectroscopy. Electrons dress the phonon quasi-particles and, thus, modify phonon frequencies, decrease their life-times and, ultimately, causes corresponding spectral broadening \cite{hasdeo2016fermi}. These effects in the Raman spectra were shown to be dependent on the materials Fermi level ($E_\mathrm{F}$), because the electron-phonon coupling can be changed by suppressing some Raman scattering paths due to the Pauli exclusion principle, thereby, offering a unique way to monitor accurately $E_\mathrm{F}$ in the family of $sp^2$ carbon nanostructures \cite{dresselhaus2010characterizing}, including the influence of different environments, such as different substrates \cite{allard2010graphene,forster2013dielectric,wang20172d}.

Twisted bilayer graphene (TBG) represents a new class of $sp^2$ carbon nanostructures for the development of graphene-based twistronics, with novel properties, such as (i) twist-angle $\theta$-dependent van Hove singularities in the electronic density of states (DOS) for large-angle TBGs ($2^{\circ} < \theta < 30^{\circ}$) \cite{li2010observation,jorio2013raman}; (ii) unusual superconductivity at the magic-angle ($\theta \sim 1.1^{\circ}$) \cite{cao2018unconventional}; and (iii) lattice reconstruction with the formation of strain soliton and topological points for low-angle TBGs ($\theta < 1.1^{\circ}$) \cite{yoo2019atomic,gargiulo2017structural,gadelha2021localization}. Recent spectroscopic and imaging resolved with nanometer spatial resolution (nano-Raman) measurements show that, in the low-angle regime, the electronic and phononic structures feature significantly localized properties \cite{gadelha2021localization}. Consequently, the C--C stretching energy uncertainty, as measured by the Raman G-band phonon linewidth ($\Gamma_{\rm{G}}$), varies spatially, following the moiré lattice structure governed by atomic reconstruction. Additionally, it was shown, at the micro-metric scale, that $\Gamma_{\rm{G}}$ reaches a maximum at the magic-angle, decreasing towards the expected graphene linewidth far below and above $1^{\circ}$, both in the large- and low-angle domains. These findings motivate exploring the Kohn anomaly in TBGs using Raman spectroscopy gate-doping experiments. These experiments enable the quantification of the importance of electron-phonon in $\Gamma_{\rm{G}}$, as contrasted to other possible structural effects \cite{cocemasov2013phonons,lamparski2020soliton,gadelha2021nano}, thus shedding light into the role of phonons in TBG strongly-correlated phenomena \cite{einenkel2011possibility,profeta2012phonon,wu2018theory,lian2019twisted,wu2019phonon,lewandowski2021pairing,choi2021dichotomy}. 

\begin{figure}[!hbtp]
\centering
\centerline{\includegraphics[width=80mm]{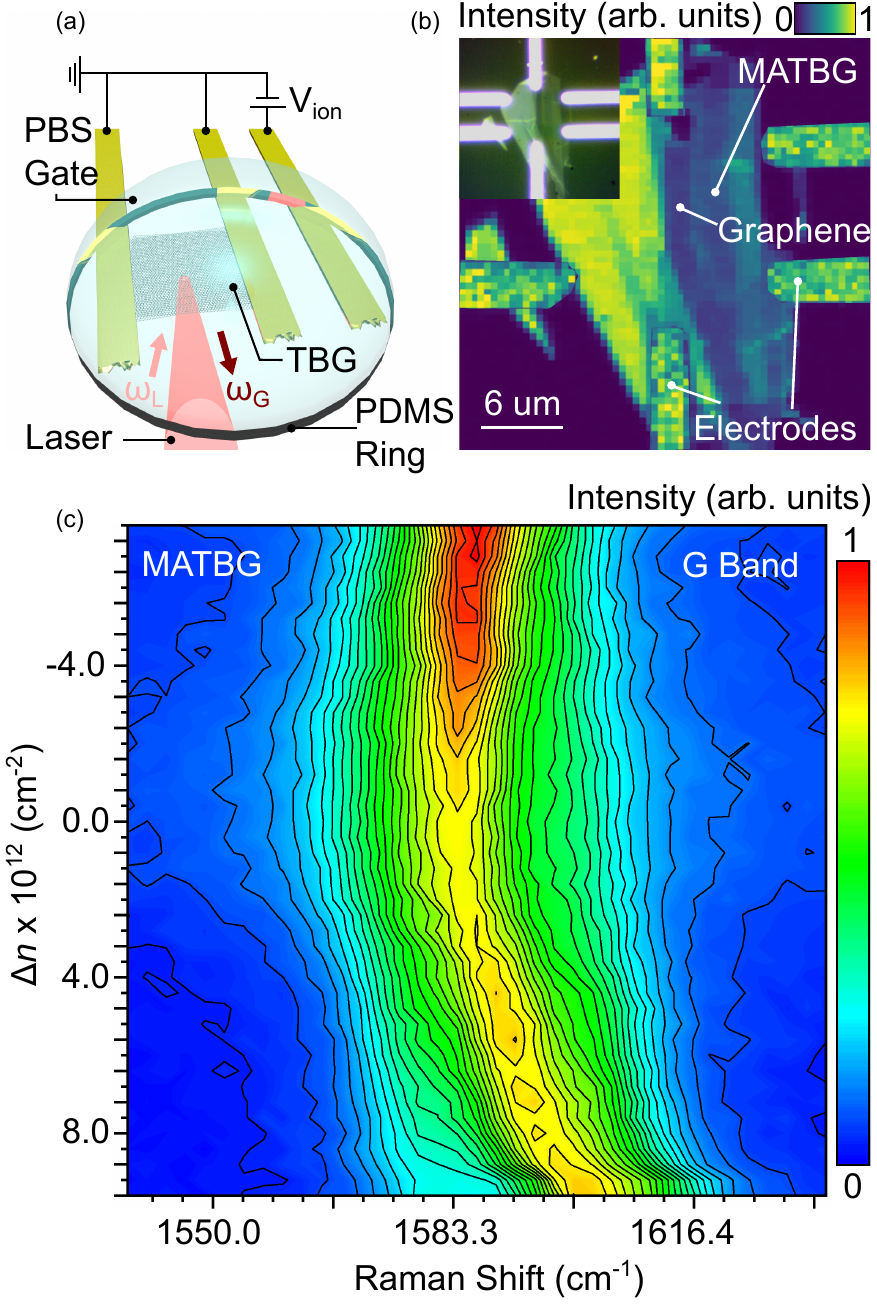}}
\caption{(a) Schematic of the TBG electrochemical device with bottom illumination and Raman detection. We exhibit the contacts (yellow stripes) with the electrical wiring, and the ionic liquid drop (transparent blue) covering the entire device. (b) Optical images of a magic-angle TBG device:  (main panel) G band Raman intensity; (inset) photo from the eyepiece of the microscope. (c) magic-angle TBG G band spectra vs. $\Delta n$. Spectra obtained with 3 accumulations of 20\,s (for Bernal and large angle TBGs, 3$\times$10\,s accumulation). We use laser powers lower than 1\,mW in the sample to avoid laser-induced sample heating. }
\label{fig:1}
\end{figure}

Figure\,\ref{fig:1} shows the schematics of these electrochemical TBG devices (a), and an optical image of a representative device (b). Electrical contacts were pre-patterned on a coverslip by optical lithography, followed by deposition of a 1\,nm chromium adhesive layer and a 50\,nm gold layer. TBG samples were prepared by a dry ``tear-and-stack'' method, using a polydimethylsiloxane (PDMS) semi-pyramidal stamp covered with a polycarbonate (PC) sheet, as described in \cite{gadelha2021twisted}. The TBGs were then transferred onto the pre-patterned electrical contacts for {\it in situ} electrical and optical studies. An electrochemical gate composed of a phosphate-buffered saline (PBS) solution is utilized (NaCl, 0.137 M; KCl, 0.0027 M; Na$_2$HPO$_4$, 0.01 M; KH$_2$PO$_4$, 0.0018 M). A potential ($V_{\rm ion}$) is applied to a free electrode that is not connected electrically to graphene, with $V_{\rm ion}$ kept low enough to avoid electrical breakdown, while the electrical contacts connected to graphene are grounded. The ionic liquid behavior is calibrated utilizing a Bernal bilayer graphene deposited on 285\,nm thick SiO$_2$, which has a known capacitance \cite{PhysRevLett.105.166601}. 

Micro-Raman scattering experiments were performed on an inverted microscope, with an oil-immersion objective ($N.A. = 1.4$), excited with a HeNe laser ($\lambda = 633$\,nm). The back-scattered light is collected by an spectrometer equipped with a charge-coupled device (CCD) and a 600\,g/l grating (Andor Solis). Figure\,\ref{fig:1}(c) shows the spectral evolution of the magic-angle TBG Raman G band spectra as a function of the carrier concentration ($\Delta n$).

\begin{figure}[!hbtp]
\centering
\centerline{\includegraphics[width=86mm]{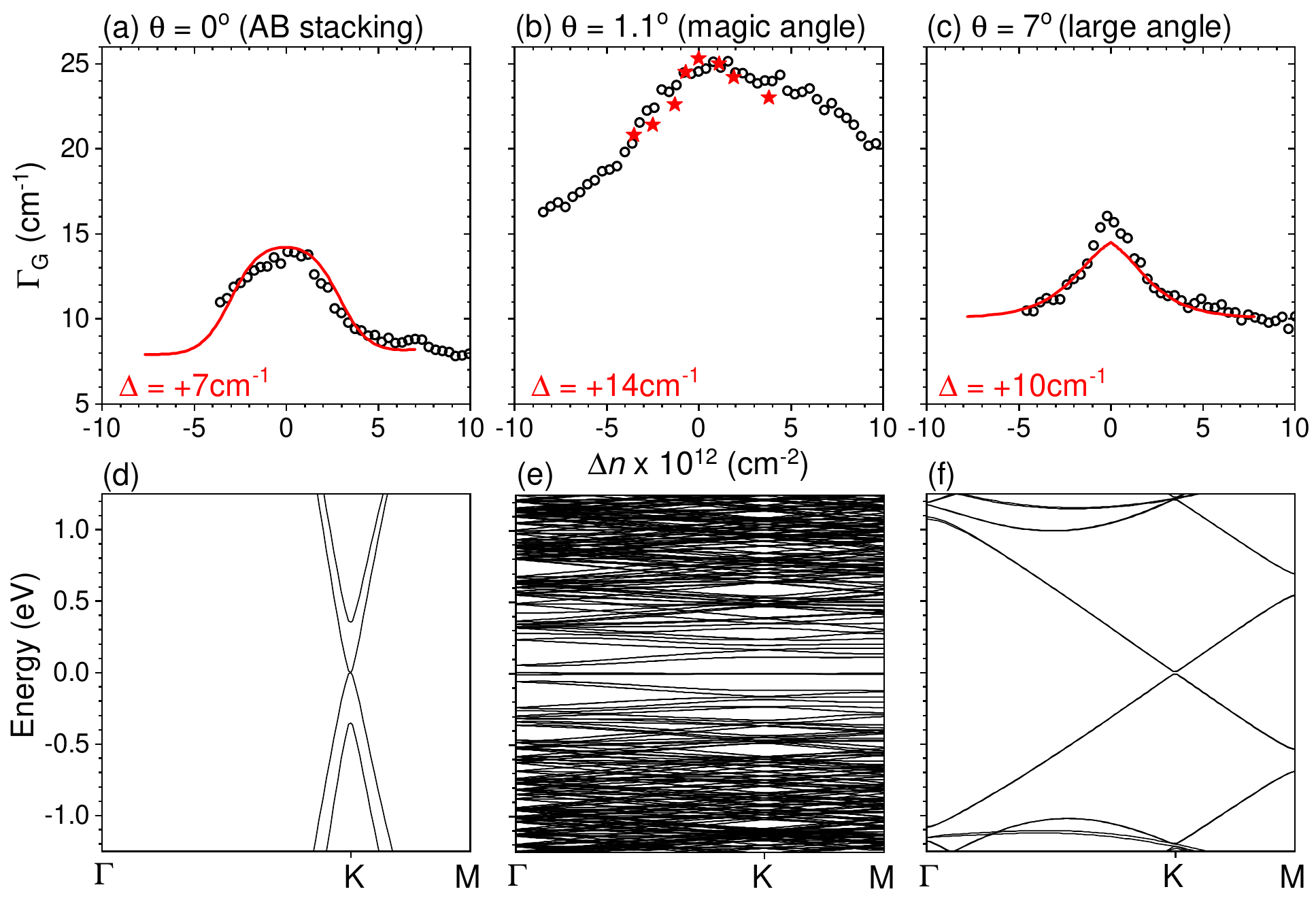}}
\caption{(a-c) gate dependent G-band full-width at half-maximum, $\Gamma_{\rm G}$, and (d-f) electronic bandstructures for Bernal (a,d -- $\theta=0^{\circ}$), magic-angle (b,e -- 1.1$^{\circ}$) and large angle (c,f -- $7^{\circ}$) TBGs. In the a-c panels, dots are experimental data while red lines and stars are theoretical predictions, up-shifted according to the $\Delta$ value displayed in each panel.}
\label{fig:2}
\end{figure}

Figs.\,\ref{fig:2}(a-c) display the gate-dependent G-band full-width at half-maximum ($\Gamma_{\rm G}$), for (a) AB (Bernal) stacked bilayer graphene ($\theta=0^{\circ}$), (b) magic-angle TBG ($\theta \sim 1.1^{\circ}$) and (c) large-angle TBG ($\theta \sim 7^{\circ}$). We fit the G bands with a single Lorentzian to extract $\Gamma_{\rm G}$, and the band linewidths values reported here are shifted by $-2$\,cm$^{-1}$ from the measured values to account of the spectrometer resolution, measured with an Ar/Ne lamp. Although gate experiments change the intensities, frequencies, and linewidths of different Raman bands in graphene-related systems \cite{hasdeo2016fermi}, here we focus on the $\Gamma_{\rm{G}}$ behavior to collect information related to the electron-phonon coupling.  

The phonon frequency is given by $\omega_q^{\nu} = \omega_q^{(0)\nu} + \omega_q^{(2)\nu}$, where $\omega_q^{(0)\nu}$ and $\omega_q^{(2)\nu}$ are, respectively, the unperturbed phonon energy and the frequency correction due to renormalization by the electron-phonon interaction (Kohn anomaly) \cite{park2028ephinteract,hasdeo2016fermi}. The phonon linewidth is directly obtained from the imaginary part of $\omega_q^{(2)\nu}$: 
\begin{eqnarray}
    \Gamma_{q}^{ph,\nu} = \Im (\omega_q^{(2)\nu}) = 4\pi \sum_{n,m} \int \frac{dk}{A_{BZ}} |M_{ep}^{\nu}(k,k+q)|^2 \times \nonumber \\
    \left [ f(E_k^n)-f(E_{k+q}^m) \right ] \delta(\hbar \omega_q^{(0)\nu} - E^m_{k+q} + E^n_k)\,
    \label{eq:1}
\end{eqnarray}
where $E_k^n$ and $E_{k+q}^m$ are the energies for the electronic states before and after the $q$-phonon mediated scattering, $f(E_k^n)$ is the Fermi distribution, $M_{ep}^{\nu}(k,k+q)$ is the electron-phonon matrix element, and $A_{\rm BZ}$ is the area of the first Brillouin zone where the $k$-integral is performed. 

Generally, previous work on $sp^2$ carbon systems \cite{bonini2007,pisana2007,lazzeri2006nonadiabatic,yan2008observation,das2009phonon,mafra2012characterizing,mafra2009observation,tsang2007doping,farhat2007phonon,dresselhaus2010characterizing,lazzeri2006phonon} considered $M_{ep}^{\nu}(k,k+q)$ as weakly varying (almost constant over the experimental range), and the phonon behavior (frequency and linewidth) can be reduced to the calculation of the joint density of states (JDOS), while taking the Pauli exclusion principle into account: $\Gamma_{\rm G}^{ph} \propto \textrm{JDOS} (\hbar\omega_G) \left [ f(-\hbar\omega_G /2) - f(\hbar\omega_G /2) \right ]$, where $\omega_G$ is the frequency of G band phonons. This simplified expression reproduces the behavior of single-layer, AB-bilayer, and large-angle TBG, but it fails when applied to magic-angle TBG, as shown in this work. In the former cases, the simple electronic structures, with linear or parabolic energy dispersions, as illustrated in Fig.\,\ref{fig:2}(d,f), are obtained in a large range of energies around the $E_\mathrm{F}$ and, therefore, their low-energy bands can be accurately modeled using a single-particle tight-binding model, even when the system is doped. However, around and below the magic-angle TBGs, strong electron localization occurs near the $E_\mathrm{F}$ \cite{nguyen2021electronic}, therefore leading to strong Coulomb interactions \cite{Kerelsky2019}. These Coulomb interactions have been shown to smooth out the spatial distribution of charge carriers and to lead to significant changes in the electronic structure upon doping \cite{Tommaso2019,Louk2019,Tommaso2020,Goodwin_2020,Lewandowski2021,Ceae2107874118}. Due to these doping effects on the electronic structure and the effects of atomic reconstruction on phonon properties in small-angle TBG \cite{lamparski2020soliton}, the doping dependence of electron-phonon interactions in magic-angle TBG cannot be described simply as in the cases of non-twisted systems and large angle TBG, as detailed below. 

In order to take into account the doping effects, we first use an adjusted electronic model, \textit{i.e.}, a single-particle tight-binding Hamiltonian modified by adding a Hartree potential term \cite{Tommaso2020,Goodwin_2020}, that depends on doping concentration \cite{SI}. We then perform atomistic calculations of the frequency $\omega_q^{(2)\nu}$ in Eq.(1), with the electron-phonon matrix elements for the G band defined as \cite{gunst2016first,choi2018strong}
\begin{eqnarray}
M_{ep}^{G}(k,n,m) = \sqrt{\frac{\hbar}{2M_C \omega^0_G}} \sum_{i \alpha} {e^G_{i \alpha} \bra{m,k} \partial_{i \alpha} {\hat H} \ket{n,k}} 
\label{eq:2}
\end{eqnarray}
where $M_C$ is the mass of a carbon atom, $\partial_{i \alpha} {\hat H}$ is the derivative of the tight-binding Hamiltonian with respect to the position of the \textit{i}-th atom along the $\alpha$-axis, $\ket{n,k}$ are the eigenfunctions of $\hat H$ with eigenvalues $E^n_k$, and  the polarization vectors $e^G_{\kappa\alpha}$ come from phonon calculations \cite{lamparski2020soliton}. The results obtained using Eq.~\ref{eq:1} with the matrix elements from Eq.~\ref{eq:2} are included in Fig.~\ref{fig:2} (red lines and stars in the (a-c) panels). We include few stars in Fig.~\ref{fig:2}(b) because the model utilized here is limited to low doping. The calculated results are offset from the experimental values by $\Delta$ (see values in the figure). For Bernal bilayer graphene  ($\theta = 0^{\circ}$), $\Delta = +7$\,cm$^{-1}$, consistent with values reported in the literature and attributed to inhomogeneous broadening and non-electron-phonon coupling effects (anharmonicity) \cite{das2009phonon,mafra2012characterizing,hasdeo2016fermi}. For the other TBGs, the values of $\Delta$ are higher, with $\Delta = +10$\,cm$^{-1}$ for large angle TBG ($\theta \sim 7^{\circ}$) and $\Delta = +14$\,cm$^{-1}$ for magic-angle TBG ($\theta \sim 1.1^{\circ}$). Inhomogeneity, mistacking, reconstruction, strain and device quality are factors that cannot be ruled out to explain the measured $\Delta$ for the magic-angle TBG. The striking result is the considerably broader and asymmetric $\Gamma_{\rm G}$-dependence on $\Delta n$ for magic-angle TBG, as compared to AB bilayer and large angle TBG. 

\begin{figure}[!hbtp]
	\centering
	\centerline{\includegraphics[width=86mm]{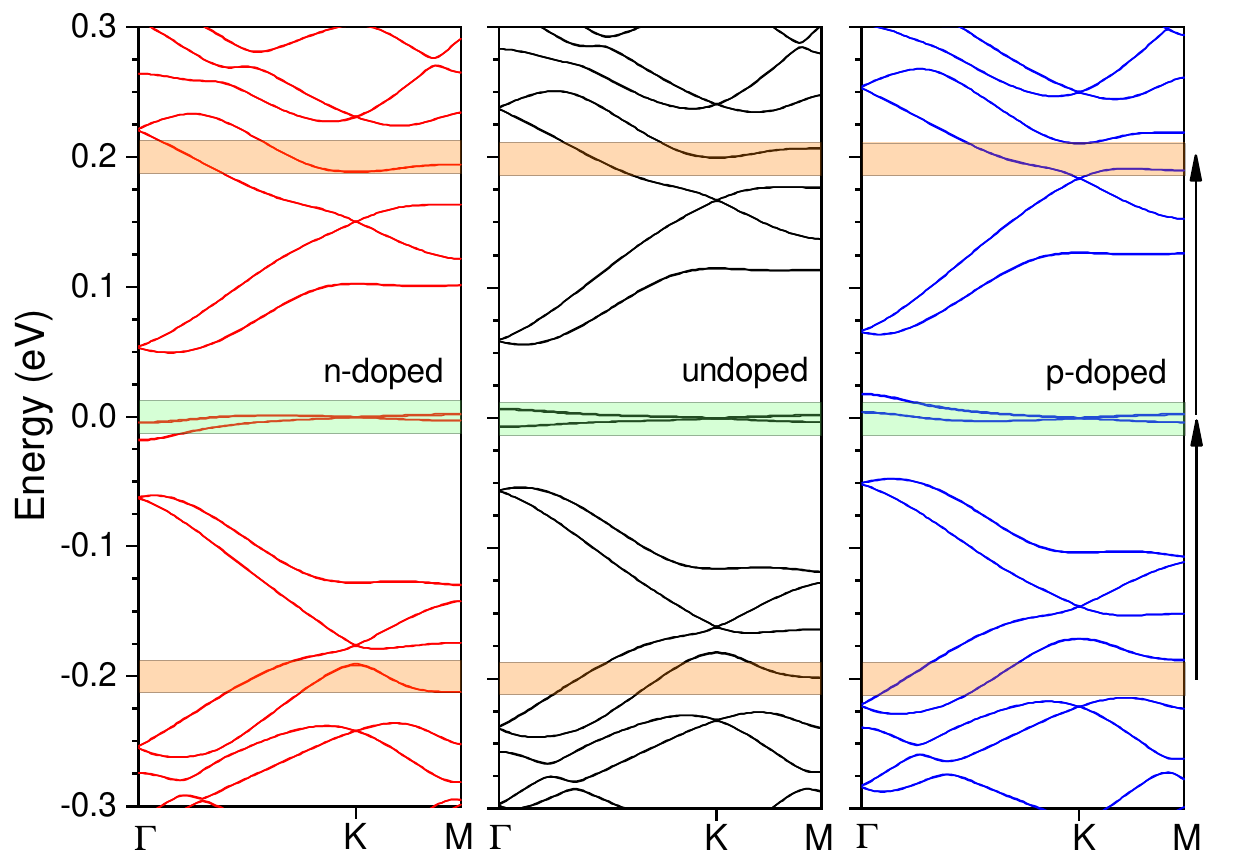}}
	\caption{Doping effect on the electronic structure of magic-angle twisted bilayer graphene: (left) n-doped, (center) undoped [zoomed in from Fig.\,\ref{fig:2}(e)], and (right) p-doped. The arrows indicate the G band energy (197\,meV), enabling the connection between the flat bands near the neutrality point (highlighted by the transparent green stripe) and valence/conduction bands (transparent orange stripes), depending on doping, as dictated by Pauli's exclusion principle.
	}
	\label{fig:3}
\end{figure}

Our results confirm that the electron-phonon coupling strength $\lambda$ is larger for magic-angle TBG, when compared to the other structures in Fig.\,\ref{fig:2}, as predicted theoretically \cite{choi2018strong}. The differences between the TBGs electron-phonon coupling strengths are direct consequences of their electronic bandstructures (see Figs.\,\ref{fig:2}(d-f)). The broader dependence on $\Delta n$ is due to the flat band near the neutrality point in magic-angle TBGs, which requires a considerable amount of carriers to be filled or emptied, associated with the presence of other electronic bands within the G band phonon energy range  (see Figs.\,\ref{fig:2}(d-f) and Fig.\,\ref{fig:3}). Although the results here are limited to the Raman active in-plane optical phonon at $\Gamma$ (G band), calculations indicate that the largest contribution to the total coupling strength comes from the in-plane optical phonon modes near $\omega_{\rm D} = 167$ and $\omega_{\rm G} = 197$\,meV, with momenta corresponding to $q = K,K'$ and $\Gamma$, respectively. Together, they contribute to about 70\% of $\lambda$ ($\lambda_{\rm D} = 0.914$ and $\lambda_{\rm G} = 0.648$) \cite{choi2021dichotomy}. These results also suggest that the electron-phonon coupling is strong ($\lambda > 1$) for TBG and twisted tri-layer graphene (TTG) at their magic-angles, but it is an order of magnitude weaker for twisted double bilayer graphene (TDBG) and twisted monolayer-bilayer graphene (TMBG), since Bernal-stacked layers in TDBG and TMBG induce sublattice polarization in at-band states, suppressing inter-sublattice electron-phonon matrix elements \cite{choi2021dichotomy}. 

The electron-phonon coupling depends more strongly on hole doping than on electron doping \cite{choi2018strong}, which is a direct manifestation of the electron-hole asymmetry of low-energy bands in the relaxed magic-angle TBG. However, our experiments find an opposite electron-hole asymmetry behavior, \textit{i.e.}, the electron-phonon interaction is larger in the n-doped regime than in the p-doped case, as also confirmed by our atomistic simulations. In contrast to previous results \cite{choi2018strong}, in this work we include the smoothing of charge inhomogeneities due to strong Coulomb interactions in magic-angle TBG. Besides the charge inhomogeneity smoothing in the moiré super-cell, which leads to changes in the low-energy flat bands, significant changes are also obtained on the high-energy bands when the system is doped. First, as illustrated in Fig.\,\ref{fig:3}, these changes in the electronic structure lead to a reduction of available electronic states which mediate the G-phonon scatterings, thus explaining the reduction of phonon linewidth when magic-angle TBG is doped. Second, the phonon linewidth reduction is observed to be stronger in the p-doped regime than in the n-doped regime (see Fig.\,\ref{fig:3}), in agreement with the experimentally observed electron-hole asymmetry [Fig.\,\ref{fig:2}(b)].

We forsaken explain the sharp peak (kink) observed at zero doping for large-angle TBG (see Fig.\,\ref{fig:2}(c)). The $\Gamma_{\rm G}$  dependence on the $E_\mathrm{F}$ in this case, as well as for Bernal stacking, can be approximately described by the function $f(-\hbar\omega_G /2) - f(\hbar\omega_G /2)$ that smoothly varies with the $E_{\mathrm{F}}$ (including the region around $E_\mathrm{F} = 0$). In the Bernal bilayer system, $\Delta n$ scales as $\Delta n \propto E_\mathrm{F}$ because of its parabolic low energy bands. The large-angle TBGs, however, exhibits a Dirac-like linear energy dispersion, similar to that observed in monolayer graphene, and therefore $\Delta n \propto E_\mathrm{F}^2$ \cite{Kim2012} and, as a consequence, a kink at zero doping is observed for large-angle TBG. Note that the simple descriptions above are no longer valid in the case of magic-angle TBG, because its flat electronic bands and doping effects makes the model significantly more complicated (see the electronic band structures illustrated in Fig.\,\ref{fig:2}(d-f)).

In conclusion, gate-dependent Raman spectroscopy measurements and atomistic simulations in TBGs show a broad and p/n-asymmetric $\Gamma_{\rm G}$ doping behavior at the magic-angle, in clear contrast to the behavior observed for both larger and smaller twist angles. These results support previous theoretical studies that investigated the relevance of electron-phonon coupling in magic-angle TBGs in order to accurately understand strongly-correlated phenomena, such as  superconductivity and its robustness in twisted layered systems \cite{choi2018strong,choi2021dichotomy}.


\subsection*{Acknowledgments}

The authors thank Prof Leaonardo C. Campos for helpful discussions. This work was supported by CNPq (302775/2018-8), CAPES (RELAII and 88881.198744/2018-01) and FAPEMIG, Brazil. A.C.G. acknowledges partial support from DoE Award No. DE-SC0008807. V-H.N. and J.-C.C. acknowledge financial support from the F\'ed\'eration Wallonie-Bruxelles through the ARC on 3D nano-architecturing of 2D crystals (16/21-077), from the European Union’s Horizon 2020 Research Project and Innovation Program — Graphene Flagship Core3 (881603), from the Flag-Era JTC project ``TATTOOS'' (R.8010.19), and from the Belgium FNRS through the research project (T.0051.18). V.M., M.L acknowledge support from NY State Empire State Development’s Division of Science, Technology and Innovation (NYSTAR).
Computational resources have been provided by the CISM supercomputing facilities of UCLouvain and the CECI consortium funded by F.R.S.-FNRS of Belgium (2.5020.11).




\end{document}